%% file: paper.tex
\documentclass[fleqn,usenatbib]{mnras}

\usepackage{newtxtext,newtxmath}

\usepackage[T1]{fontenc}

\DeclareRobustCommand{\VAN}[3]{#2}
\let\VANthebibliography\thebibliography
\def\thebibliography{\DeclareRobustCommand{\VAN}[3]{##3}\VANthebibliography}

\usepackage{graphicx}	
\usepackage{amsmath}	
\usepackage{subfiles} 
\usepackage{orcidlink}
\usepackage{nccmath}
\newcommand{\Lgal}{\textsc{\textbf{L-galaxies}}}

\newcommand{\ltsimeq}{\raisebox{-0.6ex}{$\,\stackrel
        {\raisebox{-.2ex}{$\textstyle <$}}{\sim}\,$}}

\title[Red sequence at cosmic noon]{The role of mergers and rejuvenation in the buildup of the quiescent population at cosmic noon}

\author[J. E. Harrold et al.]
{Jimi E. Harrold\orcidlink{0000-0001-7755-6638}$^{1}$, 
Omar Almaini\orcidlink{0000-0001-9328-3991}$^{1}$,
Frazer R. Pearce\orcidlink{0000-0002-2383-9250}$^{1}$,
Robert M. Yates\orcidlink{0000-0001-9320-4958}$^{2}$, 
Dave Maltby $^{1}$, \newauthor
Kate Rowlands\orcidlink{0000-0001-7883-8434}$^{3}$,
Vivienne Wild\orcidlink{0000-0002-8956-7024}$^{4}$,
Maya Skarbinski\orcidlink{0009-0004-0844-0657}$^{3}$,
Thomas de Lisle $^{1}$
\\
$^{1}$School of Physics \& Astronomy, University of Nottingham, Nottingham, NG7 2RD, UK\\
$^{2}$Centre for Astrophysics Research, University of Hertfordshire, Hatfield, AL10 9AB, UK\\
$^{3}$AURA for ESA, Space Telescope Science Institute, 3700 San Martin Drive, Baltimore, MD 21218, USA; William H. Miller III \\
$^{4}$School of Physics and Astronomy, University of St Andrews, North Haugh, St Andrews KY16 9SS, UK\\
}

\date{Accepted XXX. Received YYY; in original form ZZZ}

\pubyear{\the\year{}}

\begin{document}
\label{firstpage}
\pagerange{\pageref{firstpage}--\pageref{lastpage}}
\maketitle

\begin{abstract}

We investigate the quenching of galaxies using a mock observational lightcone generated from the Semi-Analytic Model (SAM) L-Galaxies, closely matched to observations from the UKIDSS Ultra Deep Survey (UDS). 
The sample is used to study merging, rejuvenation, and visibility times for star-forming, quiescent, and post-starburst (PSB) galaxies, to assess the impact on the build-up of the passive galaxy mass functions.
 We find, for example, that a typical PSB ($M_\ast\sim10^{10}$\,M$_\odot$) at $z\approx1$ has a 15 per cent likelihood of merging and around a 25 per cent likelihood of rejuvenating within 1 Gyr of being identified. Applying these rates and timescales to the observational data, we estimate the fraction of quiescent galaxies that passed through a PSB phase. We find that $18 - 28$ per cent of the build-up in the massive end ($M_\ast>10^{10}$\,M$\,_\odot$) of the passive mass function at $1<z<2$ can be explained by PSBs, with the contribution declining to $\sim5$ per cent by $z \simeq 0.5$. Accounting for mergers and rejuvenation  reduces the inferred PSB contribution by approximately a factor of two. At lower stellar masses ($M_\ast < 10^{10}$\,M$_\odot$), rapid quenching through a PSB phase explains a significantly larger fraction of the growth in the passive mass function. With a visibility time of $\sim$ 0.75 Gyr,  we find that around $60-80$ per cent of low-mass passive galaxies underwent a PSB phase. Our findings provide further evidence that low- and high-mass galaxies follow different quenching pathways.
\end{abstract}



\section{Introduction}
\subfile{sections/introduction}

\section{The Data}
\subfile{sections/methods }

\section{Simulated Galaxy Sample}
\subfile{sections/mock_results}

\section{Application to observational results: the number density evolution of the quiescent population}
\subfile{sections/observation_results}

\section{Conclusions}
\subfile{sections/conclusion }

\section*{Acknowledgments}
This work was supported by the Science and Technology Facilities Council (STFC) grants ST/X000982/1 and ST/X006581/1. 
VW acknowledges the STFC grant ST/Y00275X/1 and Leverhulme Research Fellowship RF-2024-589/4.
We gracefully acknowledge support from the NASA Astrophysics Data Analysis Program (ADAP) under grant 80NSSC23K0495.
\section*{Data Availability}
The observational data presented in this work is available from public archives, further details of which can be obtained from the UDS web page (https://www.nottingham.ac.uk/astronomy/UDS/). A public release of the processed and simulated data is in preparation, and can be obtained on reasonable request to the corresponding author. 


\bibliographystyle{mnras}
\bibliography{ref} 




\appendix
\section{Merger and Rejuvenation Tables}
\subfile{sections/appendix}


\bsp	
\label{lastpage}
\end{document}

%% file: sections/introduction.tex
Local galaxy populations exhibit a clear bimodality, with most galaxies categorized as blue, star-forming discs or red, passive spheroidal systems \citep{strateva_color_2001,hogg_luminosity_2002}. Over time, galaxies transition from the blue cloud to the red sequence as star formation quenches, as evidenced by the growing passive population \citep{muzzin_evolution_2013-1,foltz_evidence_2015,balogh_evidence_2016}. 
Although the majority of galaxies would eventually exhaust their available gas and gradually stop forming stars if left isolated, many quench much earlier due to processes such as AGN feedback, supernova-driven winds, or environmental effects such ram-pressure stripping.
 While the various proposed mechanisms for quenching can all viably explain a galaxy transitioning from the blue sequence to the red sequence, the transient nature of many quenching mechanisms, combined with the lack of clear distinguishing signatures, can make isolating their relative contributions challenging.
\par

Semi-analytic models (SAMs) provide a powerful tool to explore quenching mechanisms, offering the ability to simulate large galaxy populations within cosmological volumes efficiently \citep{white_galaxy_1991,springel_gadget_2001,baugh_primer_2006,somerville_physical_2015}. SAMs provide insight into statistical trends, rare populations, and environmental effects, while enabling targeted studies on populations of recently quenched galaxies. Despite broadly reproducing observed stellar mass functions and clustering \citep{asquith_cosmic_2018,cecchi_quiescent_2019}, SAMs struggle to simultaneously match the passive fraction across redshifts, hinting at gaps in our understanding of quenching physics \citep{asquith_cosmic_2018,henriques_origin_2019,cecchi_quiescent_2019,ayromlou_comparing_2021}. Feedback from the central supermassive black hole, known as AGN feedback, is a popular mechanism employed by SAMs and typically acts to prevent star formation by shutting off the cooling of the hot gas supply of a galaxy for 0.5–5 Gyr \citep{croton_many_2006,zheng_comparison_2020}. While implementing AGN feedback in SAMs historically improved agreement with observed passive populations, SAMs still struggle with an overproduction of low-mass passive galaxies at high redshifts, potentially suggesting overly efficient quenching in such systems \citep{asquith_cosmic_2018}. Recent attempts to fix this problem, either via allowing stronger retention of hot gas in satellite and orphan systems \citep{harrold_correcting_2024}, or by allowing more efficient star-formation after mergers \citep{araya-araya_simultaneously_2025}, have successfully managed to rectify this overly efficient quenching without significantly altering the existing AGN models.
\par

The main difficulty with effectively modeling the precise nature and relative contribution of each quenching mechanism is their sometimes transient or periodic nature and often indirect disruption of star formation. This makes it challenging to identify significant quantities of actively quenching galaxies in observations for use in model calibration. A class of recently and rapidly quenched galaxies known as Post-StarBursts (PSBs) have been used to study quenching mechanisms while they are potentially still active within the host galaxy. Identified by strong Balmer absorption lines and little ongoing star formation, PSB galaxies may offer a direct window into quenching mechanisms \citep{Wild2009}. Their spectra indicate rapid quenching within the past Gyr. While in low-redshift observations, the star formation histories (SFHs) of PSBs do typically exhibit a distinctive burst and drop off in star formation, at high-redshift ($z>1$) the naturally higher rates of star formation can lead to PSB signatures without necessarily requiring a major burst before quenching.
The distinction between "true" PSBs and galaxies with continuously high SF before quenching tends not to be important for our purposes, as we are primarily interested in galaxies that underwent a rapid quenching event. Advances in photometric identification methods, such as the PCA "supercolour" (SC) approach \citep{wild_new_2014,wild_evolution_2016}, have significantly expanded PSB samples.
However, to fully understand how galaxies transition from the PSB phase to the fully quiescent red sequence requires reliable estimates of PSB visibility timescales.
\par

In theory, almost all galaxies currently undergoing a PSB phase will fully transition to the older passive population as the A-class stars causing the PSB signatures die and their distinctive imprint on the Spectral Energy Distribution (SED) fades. While PSB signatures are expected to be present for $\sim 0.5-1$ Gyr, the timescale over which these features are detectable could be different depending on how rapidly the galaxy quenches, and the method used to select them. \cite{wild_evolution_2016} showed that up to 50\% of the 
quiescent population in the redshift range $0.5<z<1.5$ can be explained 
by rapid quenching, assuming a PSB visibility time of around 500 Myr.
\cite{belli_mosfire_2019} estimated the contribution using spectroscopic data, finding that PSBs build up around 50\% of passive galaxies at $z\approx2$ and 20\% at $z\approx1.4$, with a visibility timescale of 250-500 Myr. Low-mass ($\log($M/\,M$_\odot)<10.5$) PSBs present additional challenges, as many are expected to merge or rejuvenate, complicating our attempts to model their contribution to the build-up in the passive mass function using observational data. These processes are influenced by environmental factors, with low-mass PSBs frequently found in dense environments, indicating a link to satellite infall and subsequent evolution \citep{socolovsky_enhancement_2018, wilkinson_starburst_2021, taylor_role_2023}. Accurately constraining these processes is essential for understanding the growth of the quiescent population over time.
\par

In this study, we compare the SAM \Lgal{} against observational data from the UKIDSS Ultra Deep Survey (UDS), aiming to match the observational catalogue used in the PSB study of \cite{wilkinson_starburst_2021}. 
We utilize the mock lightcone, generated in 
\cite{harrold_correcting_2024},
which is designed to closely match the UDS observational constraints (using the same observational filters, etc), and produces estimates for galaxy properties such as stellar mass in a manner consistent with observations. We use the model lightcone to investigate the influence of merging, rejuvenation, and visibility times on the build-up of the passive galaxy mass function.
\par

Throughout this paper we assume a cosmology of $\text{h}=0.7, \Omega_m=0.3, \Omega_\Lambda=0.7$. \Lgal{} was run on the rescaled $480\text{Mpc}^{3} \text{h}^{-3}$ Millennium simulation using parameters outlined in \cite{henriques_galaxy_2015}, and property values have been rescaled to our assumed cosmology in post-processing for all plots and analysis presented in this work. 
\par

%% file: sections/methods.tex
\subsection{The UDS galaxy sample}
\begin{figure}
    \includegraphics[width=\columnwidth]{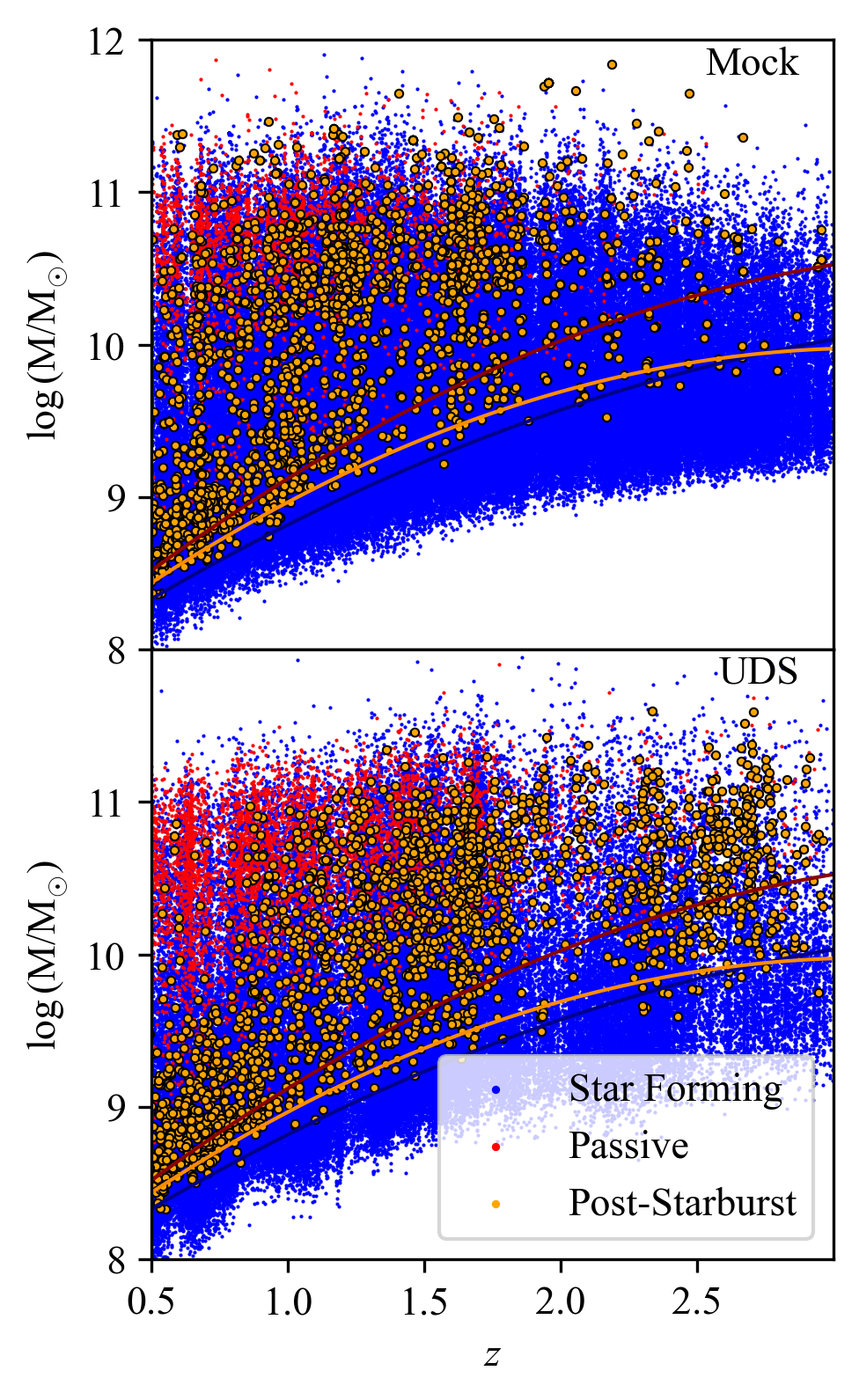}
   \caption{Galaxies from the Mock lightcone (top panel, generated using \Lgal{}), compared to galaxies from the UDS catalogue (lower panel) \citep{wilkinson_starburst_2021}. Observationally derived stellar masses and redshifts are shown in both panels. Galaxies are color-coded by type: blue for star-forming, red for passive, and orange for post-starburst (PSB). Coloured solid lines indicate the 95\% stellar mass completeness limits for each galaxy type as a function of redshift. Notable differences emerge between the Mock and UDS distributions, particularly at higher redshifts. For instance, the Mock contains significantly fewer galaxies with masses above $10^{10.5} $\,M$_\odot$ at $z > 2$ compared to the UDS and significantly more low-mass PSB galaxies.}
    \label{fig:mock_uds_mass_redshift}
\end{figure}

The observational catalogue we use is described in \cite{wilkinson_starburst_2021}, which is based on a $K$-band selected catalogue from the UKIDSS UDS DR11 release. The UDS covers 0.77 square degrees in the $J=25.6$, $H=25.1$ and $K=25.3$ bands. Additional imaging of the UDS field has been taken over a wide range in wavelength. The catalogue of \cite{wilkinson_starburst_2021} combines the UKIDSS near-infrared imaging with deep optical imaging in $B, V, R, i', z'$ from  Subaru Suprime-CAM, $u'$ imaging from the  Canada-France-Hawaii Telescope MegaCam instrument, $Y$-band imaging from VISTA, and near/mid-IR imaging from Spitzer IRAC at  $3.6\mu m $ and $4.5\mu m$. The overlapping area with 12-band imaging is 0.62 square degrees, allowing for the masking of artefacts and  bright stars. Further details of these data and the depths achieved can be found in \cite{almaini_massive_2017} and \cite{wilkinson_starburst_2021}.


\subsection{Lightcone construction}
The lightcone, which is briefly described in \cite{harrold_correcting_2024}, is constructed using the Semi-Analytic Model (SAM) \Lgal{} \citep{springel_simulations_2005,guo_dwarf_2011,henriques_galaxy_2015,henriques_l-galaxies_2020}. This model produced a better match to the observed evolution of the star-forming and passive mass functions compared to other SAMs in the comparison study of 
\cite{asquith_cosmic_2018}, and closely matches the pair fractions observed in the UDS catalogue \citep{mundy_consistent_2017}. Specifically, we use the 2015 version of \Lgal{} \citep{henriques_galaxy_2015}, which incorporates dust attenuation as prescribed by \cite{charlot_simple_2000} and assumes a redshift-dependent dust-to-metal ratio. The lightcone spans a 1 square degree field of view over a redshift range of $0.5 < z < 3.0$, and is designed to closely replicate the depth and coverage of the UDS. In this redshift range, the time between simulation snapshots within \Lgal{} when using the Millennium-I merger trees, varies from $\sim350$ Myr at $z=0$ to around $\sim220$ Myr at $z=3$.
\par

Photometric properties of galaxies within \Lgal{} are constructed by first using the \cite{maraston_evolutionary_2005} stellar population synthesis model. Dust extinction follows the \cite{de_lucia_hierarchical_2007} method by separating dust extinction between the inter stellar medium  \citep[which uses the prescription of][]{devriendt_galaxy_1999}, and molecular birth clouds where star formation begins 
\citep[using the prescription of][]{charlot_simple_2000}. An explicit dust production and destruction model is incorporated into later versions of L-Galaxies \citep{vijayan_detailed_2019,yates_impact_2024}. This will be used to explore photometric properties in more detail in future work.
\par

The construction of the lightcone begins by tessellating the \Lgal{} simulation volume in 3D space to fully populate the intended observational region. At each simulation snapshot, galaxies are tested for inclusion within the lightcone volume. Those that fall within the volume are added to the output catalogue, and their complete evolutionary histories — spanning previous and future snapshots — are saved separately to allow for accurate reconstruction of their properties over time. After determining the position of each galaxy in the lightcone, its flux in all observational bands is recalculated accordingly.
\par

Throughout this paper, we refer to the processed output of \Lgal{}, passed through the lightcone and subjected to the same observational limits as the UDS, as the "Mock". The unfiltered raw output of \Lgal{}, without any observational constraints, is referred to simply as "\Lgal{}".
\par

\subsection{Photometric redshifts, galaxy classification, and stellar mass estimation}\label{sec:supercolors}
Photometric redshifts for both the Mock and UDS datasets were calculated using the method described in \cite{simpson_prevalence_2013} with the EAzY code \citep{brammer_eazy_2008}. This approach employs the default EAzY setup, incorporating 12 Flexible Stellar Population Synthesis (FSPS) SED templates \citep{conroy_propagation_2010} augmented with three simple stellar population (SSP) models. These SSP models, with ages of 20, 50, and 150 Myr, assume a Chabrier IMF and 0.2 solar metallicity, enhancing the representation of starburst galaxies and complementing the FSPS templates. This setup minimizes scatter and outlier fractions relative to spectroscopic redshifts in the UDS, with varying metallicities having negligible effects. Using around 8,000 objects in the UDS with secure spectroscopic redshifts, we achieve a normalized median absolute deviation (NMAD) of $\sigma_{\text{NMAD}}=0.019$ and an outlier fraction ($|\Delta z|/[1+z] > 0.15$) of $\sim 3$\%. Further details on the method are available in \cite{simpson_prevalence_2013} and \cite{wilkinson_starburst_2021}.
\par

Galaxy classifications in both the UDS and Mock datasets were determined using the Principal Component Analysis (PCA)-based Super-Colour (SC) technique described in \cite{wild_new_2014}, and applied to the UDS DR11 catalogue in \cite{wilkinson_starburst_2021}. The PCA method identifies three eigenvectors (SC1, SC2, and SC3) that capture over 99.7\% of the observed variance in galaxy (photometric) SEDs. These SCs correlate with physical galaxy properties: SC1 is strongly related to mean stellar age and with dust content (higher SC1 indicates younger stellar populations), SC2 reflects the fraction of stellar mass formed in the last Gyr, and helps to break the degeneracy between dust and age (higher SC2 corresponds to recent star formation), and SC3 resolves degeneracies between stellar mass fraction and metallicity. Galaxies are classified into star-forming (SF), passive, post-starburst (PSB), and dusty categories based on their location in SC1-SC2 space, with spectroscopic studies demonstrating PSB identification accuracy of up to 80\% using this method \citep{maltby_identification_2016,wild_evolution_2016}.
\par

As described in \cite{wild_new_2014}, galaxy stellar masses are estimated through Bayesian analysis of 44,000 population synthesis models from \cite{bruzual_stellar_2003}, which span a wide range of star formation histories, dust contents, and metallicities. By fitting the first three SCs to these models, we derive probability density functions (PDFs) for physical properties such as the stellar mass and star formation rate of a galaxy, with the median value of the PDF being taken as the value for each physical property. This approach reproduces \Lgal{} stellar masses with high fidelity, with 96.1\% of Mock galaxies having SC-derived stellar masses within two standard deviations of their SAM-derived masses.
\par

The stellar mass completeness limits used were calculated in \cite{wilkinson_starburst_2021} using the method described in \cite{pozzetti_zcosmos_2010}. When comparing the mock lightcone to observations, we impose an additional minimum stellar mass completeness limit of $\log($M/\,M$_\odot)=9.5$ to both the UDS and the mock, reflecting the lowest-mass galaxies that are well resolved in \Lgal{} when run on merger trees from the Millennium simulation \citep{henriques_galaxy_2015}. The overall galaxy population for both the mock lightcone and the UDS data can be seen in Figure \ref{fig:mock_uds_mass_redshift}, which shows the SC derived masses and galaxy classifications across the entire available redshift range of $0.5<z<3$.
\par

\begin{figure*}
    \includegraphics[width=2\columnwidth]{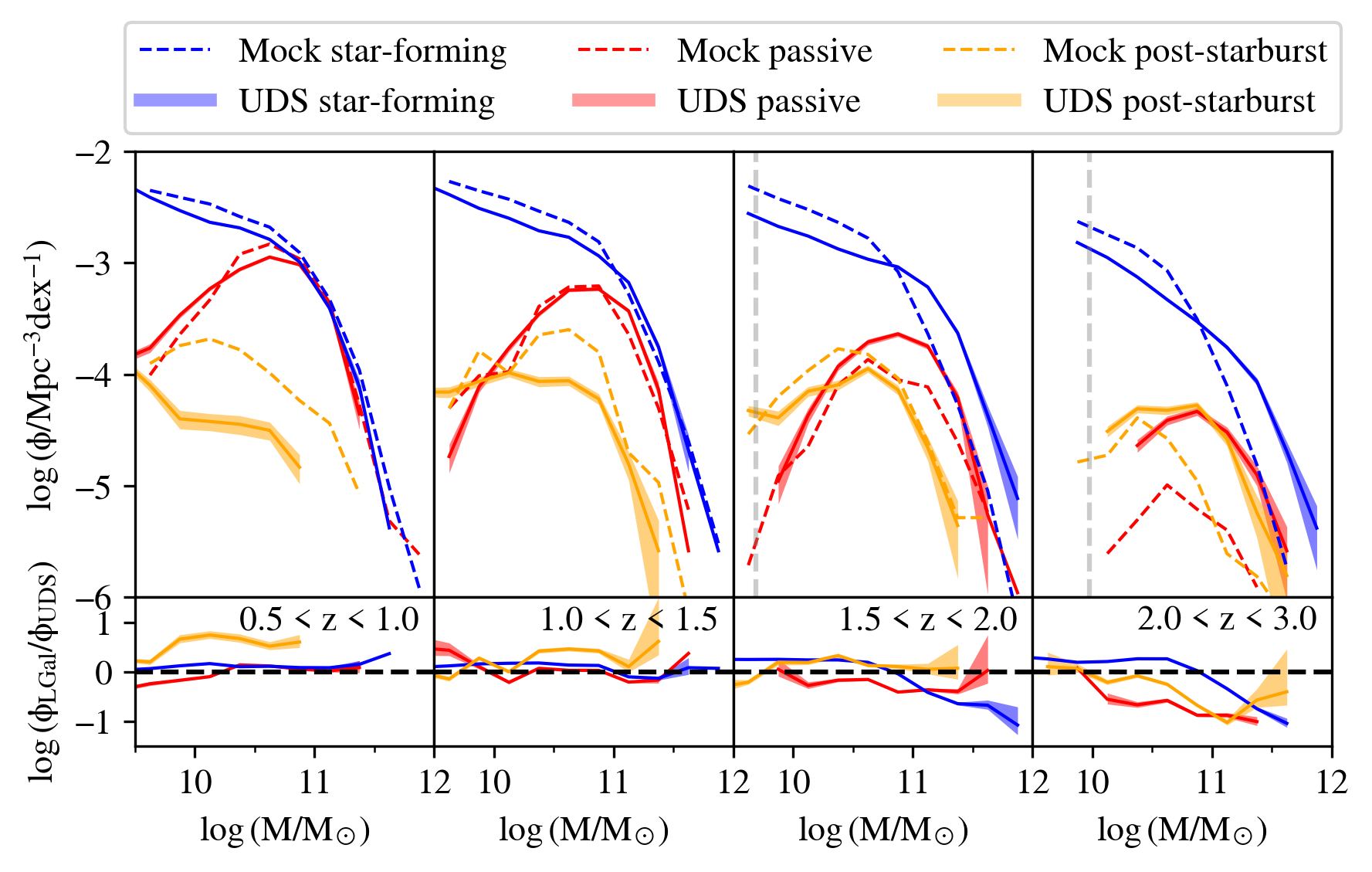}
    \caption{The SMFs from the Mock (dashed lines) and the UDS (solid lines) across the specified redshift ranges (upper panels). Blue, red, and orange lines represent star-forming, passive, and post-starburst (PSB) galaxies, respectively. Shaded regions indicate the 1$\sigma$ uncertainty on the UDS SMFs. The lower panels display the differences between the SMFs of the Mock and UDS. The SMFs include the orphan correction described in Section \ref{sec:orphan}. At lower redshifts, the Mock reproduces the UDS trends well for star-forming and passive galaxies but overpredicts the number of PSBs. At higher redshifts, however, significant deviations between the two datasets become evident.}
    \label{fig:mock_uds_smf}
\end{figure*}

\subsection{Orphan galaxies}
\label{sec:orphan}
Orphan galaxies are a subset of galaxies in SAMs whose underlying dark matter subhalo can no longer be detected by the subhalo finder. Different SAMs have historically employed varying strategies to handle orphan galaxies. Some models disrupt orphans immediately, transferring their gas and stars to the central galaxy \citep{croton_semi-analytic_2016}. More commonly, SAMs strip the orphan's hot gas reservoir and assign it a merger timescale for eventual coalescence with the central galaxy \citep{gonzalez-perez_how_2014,cattaneo_galics_2020,henriques_galaxy_2015,cora_semi-analytic_2018,lagos_quenching_2023}. Both approaches are motivated by the understanding that, without a dark-matter subhalo, the galaxy's potential well becomes too shallow to retain gas against external forces such as ram-pressure stripping. This stripping occurs on short dynamical timescales, while the orphan galaxy’s trajectory continues to be remembered until its stellar component and any remaining gas merges with the central galaxy \citep{mccarthy_ram_2008}.
\par 

Quenching via the orphan route occurs rapidly, causing orphans to be quickly classified as quiescent. In the local universe, this mechanism effectively reproduces the observed passive fraction for galaxies down to moderate masses. However, as observations have probed deeper, capturing more complete SMFs at lower masses, the discrepancies between the observed passive galaxy counts and the SAM predictions have grown \citep{asquith_cosmic_2018,donnari_quenched_2021}. \cite{harrold_correcting_2024} identified the orphan population as the primary driver of these discrepancies. To address this, we apply the retroactive correction proposed in \cite{harrold_correcting_2024}, which uses an orphan-satellite matching method to prescribe orphans with the properties of a satellite galaxy with matching properties pre-infall. Implementing this fix reduces the contribution of orphans to the low-mass passive SMF tail to approximately one third of its original level. This correction improves the representation of low-mass passive galaxies at high redshifts, aligning model predictions more closely with observations.
\par

\subsection{Stellar Mass Functions}
Figure \ref{fig:mock_uds_smf} compares the stellar mass functions (SMFs) from the UDS and the Mock lightcone, after applying the orphan adjustment described in Section \ref{sec:orphan}. One noteworthy discrepancy occurs at low redshifts, where the Mock overproduces PSB galaxies by a factor of three at $M_\ast=10^{10} $\,M$_\odot$ compared to the UDS. At redshifts above $z=1.5$  additional mismatches in the high-mass end are observed that cannot be fully attributed to cosmic variance. It is likely these arise from insufficient modeling of mass assembly downsizing in SAMs \citep{asquith_cosmic_2018}. Although, these may in part arise from differences in the $\Lambda$CDM parameters used in the Millennium-I simulation versus those adopted for UDS DR11. Although \Lgal{} employs the remapping method from \cite{angulo_one_2010} to align cosmologies, this technique provides an approximation rather than an exact match. The remapped cosmology corresponds to the standard Planck cosmological parameters: $H_0=67.3\, \rm{km\,s^{-1}}\,\rm{Mpc^{-1}}$, $\Omega_\Lambda=0.685$, $\Omega_m=0.315$, and $\sigma_8 = 0.829$.
\par

%% file: sections/mock_results.tex
\subsection{Merger and Rejuvenation Rates in \Lgal{}}\label{sec:merger_rejuvination_rates}
While the historic merger rates of galaxies undergoing a PSB phase have been studied extensively, the predicted future merger likelihood of PSB galaxies is less constrained. Previous attempts at accounting for the merger rates of PSB galaxies such as \cite{belli_mosfire_2019} have used estimates of the number of major mergers over the entire galaxy population. 
Such techniques may not account for the strong bias for certain galaxy classes, e.g. 
low-mass PSBs, which are found to reside in high-density environments on average, particularly at $z<1$
\citep[e.g.,][]{socolovsky_enhancement_2018, taylor_role_2023}. One of the key formation mechanisms for low-mass PSBs at $z<1$ is thought to be ram-pressure stripping via infall into over-dense environments \citep{socolovsky_enhancement_2018}. In addition there is the possibility that the severity of the quenching may be related to the angle of infall, with galaxies plummeting directly towards the central galaxy being more likely to quench rapidly
\citep[e.g.,][]{Jaffe2018}, and 
more likely to merge quickly than galaxies that remain on the outskirts of the group for long periods. These processes may greatly increase the likelihood of any given PSB galaxy undergoing a merger compared to their star-forming counterparts. As low-mass galaxies are more likely to be the minor galaxy during a merger event, and thus are always guaranteed to be eradicated from the PSB or quiescent population once merged, 
it stands to reason that a significant fraction of the current low-mass PSB population may soon merge out of existence.
\par

\subsubsection{Merger Rates}
Estimating global merger rates, either analytically or directly from observations, has historically been difficult. Isolating the rates for a given sub-population of galaxies, such as PSBs, compounds this difficulty. We opt to utilize the benefits of having a mock observational lightcone backed by a fully tracked simulation to generate an empirical estimate of merger rates. The merger timescale for satellite galaxies within \Lgal{} can simply be estimated by counting the time between when a galaxy first becomes a satellite and when it merges. For orphan galaxies, \Lgal{} estimates a merger time using the formula outlined in \cite{binney_galactic_1987}:
\begin{equation}
  t_{\rm friction}=\alpha_{\rm friction}\frac{V_{200c}r^2_{sat}}{GM_{sat}\ln(\Lambda)},
\end{equation}
where $M_{sat}$ is the mass of the satellite galaxy, $V_{200c}$ is the virial velocity of the central dark matter subhalo, $r_{sat}$ is the distance between the satellite and the central, $\ln{\Lambda} = \ln(1 + M_{200c} /M_{sat})$ is the Coulomb logarithm
and $\alpha_{friction}$ is the coefficient of friction (which is constrained via MCMC fitting to a value of around 2.5). While \cite{harrold_correcting_2024} discusses some of the key flaws in the overall treatment of the orphan galaxy population, these flaws exclusively relate to the baryonic properties of these galaxies, not their merger timescales. As the overall galaxy population within \Lgal{} is matched well to the observed galaxy distribution, the measured merger rates should provide a reasonable estimation of the typical merger timescales for orphan galaxies in different environments. We suggest that the probability that any given galaxy in a specific population will merge as a function of time, $t$, is directly proportional to the total fraction of simulated galaxies taken from that same population that have merged by $t$. We split the galaxies into populations based on their mass, redshift, and SC classification. We have explicitly not binned by measures such as environment due to the difficulty in accurately measuring an observed galaxy's environment and ensuring it reliably matches the measure used for the simulated galaxies. 

\begin{figure}
  \includegraphics[width=\columnwidth]{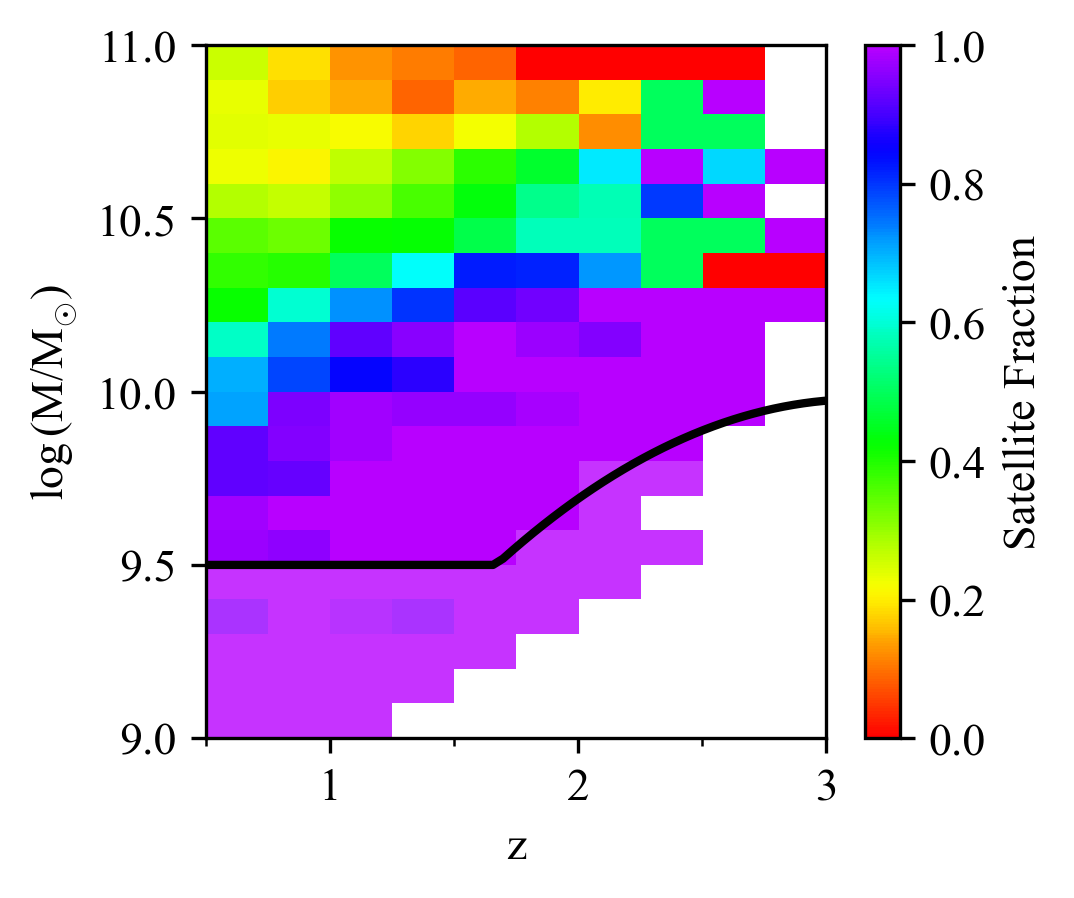}
  \caption{The fraction of PSB galaxies that are satellites in the \Lgal{} mock dataset, as a function of redshift and stellar mass. The solid black line describes the 95\% mass completeness limit for PSB galaxies in the UDS, truncated at
  $10^{9.5}$\,M$_\odot$ to represent the lower limit for resolving galaxies effectively in \Lgal{}.
  The majority of low-mass PSB galaxies are satellites at $z\ltsimeq 1.5$, with the high-redshift, high-mass population being dominated by central galaxies. Similar evidence for a low-mass satellite population has been found in the observational data (e.g. 
  \protect\citealt{socolovsky_enhancement_2018},
  \protect\citealt{maltby_structure_2018}, \protect \citealt{wilkinson_starburst_2021})}
  \label{fig:psb_satellite_fraction}
\end{figure}

\begin{figure*}
  \includegraphics[width=2\columnwidth]{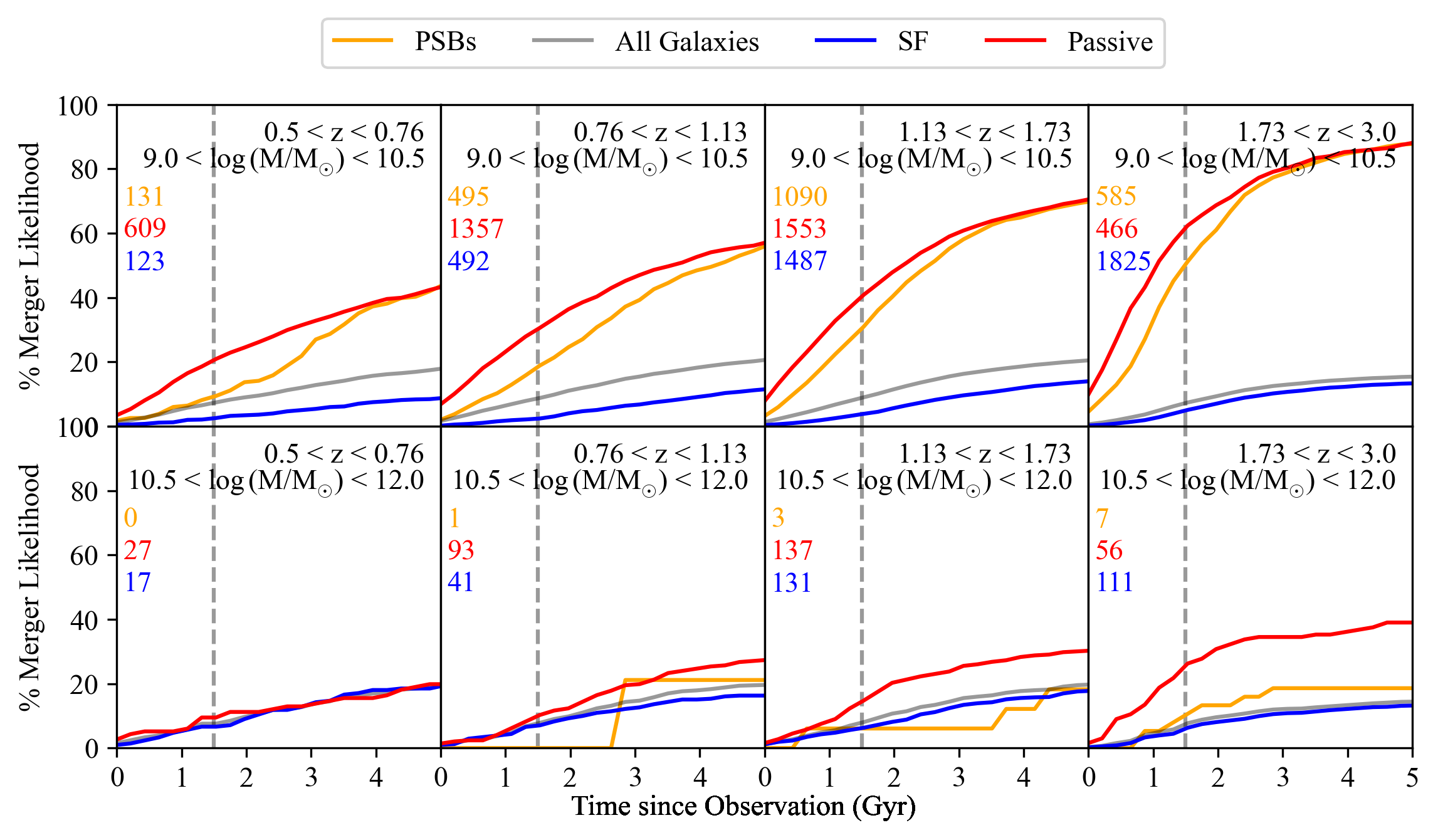}
  \caption{The measured likelihood that a galaxy will undergo a merger as a function of time since observation in the \Lgal{} mock dataset. Low mass galaxies below $10^{10.5}$\,M$_\odot$ are shown in the upper row of panels. Galaxies above $10^{10.5}$\,M$_\odot$ are shown in the bottom row. The total number of PSB galaxies, quiescent galaxies, and star-forming galaxies with valid merger rates in each bin are shown in each plot as the orange, red, and blue text respectively. The vertical dashed grey line indicates a time of 1.5 Gyrs after observations. This reflects the typical time period between the redshift bins used in this paper. Low-mass passive galaxies (red lines) and low-mass PSBs (orange lines) show significantly higher merger rates compared to a typical SF galaxy and all high-mass galaxy types. }
  \label{fig:psb_merge_rates}
\end{figure*}

The ratio of satellites to centrals for PSB galaxies is shown in Figure \ref{fig:psb_satellite_fraction}, and the cumulative distribution function of merger rates in our simulated mock for various masses and redshifts is shown in Figure \ref{fig:psb_merge_rates}. Figure \ref{fig:psb_satellite_fraction} shows that the majority of low-mass PSB galaxies are satellites, agreeing well with observations made in \cite{wilkinson_starburst_2021},
\cite{socolovsky_enhancement_2018}, and \cite{taylor_role_2023}, which showed that low-mass PSB galaxies at lower redshifts tend to exist in highly dense environments. In Figure \ref{fig:psb_merge_rates}, while low-mass PSB galaxies tend to have similar merger likelihoods to star-forming satellites within 1.5 Gyr of observation, after 1.5 Gyr they become increasingly more likely to have undergone a merger. Around 3 Gyr after observation, approximately 20\% of PSBs will have undergone a merger, compared to around 5\% of the star-forming galaxies. These values corroborate the empirical result that low-mass quiescent and PSB galaxies within \Lgal{} are on average $\sim2$ times closer to their central galaxy than equivalent low-mass star-forming galaxies.
\par

\subsubsection{Rejuvenation Rates}

Rejuvenation rates in observations are difficult to constrain. This is mainly due to the fact that most methods of inferring a galaxy's star formation history are unable to distinguish between a galaxy that has been star-forming for an extended period and one that has had an intermittent passive phase several Gyrs previously, preventing reliable detection of previously rejuvenated galaxies. In simulations, rejuvenation mechanisms for galaxies are well described and easily measured. Simulated galaxies in \Lgal{} typically rejuvenate via two key mechanisms: (i) splashbacks (galaxies which pass through the dark matter halo of a larger galaxy and become centrals again), (ii) wet mergers (the accretion of new gas brought in by merging satellites). See \cite{henriques_galaxy_2015} for details of the implementation. If we assume mock rejuvenation rates are well constrained and reasonably representative, we can use the predicted rates for a given galaxy population to attempt to account for the effects of rejuvenation on galaxies in the observational data.
\par

In this paper, we define a galaxy as having rejuvenated over a fixed timescale if at any point it is classified as quiescent and then later regains and maintains a sSFR of over $10^{-10}\, \text{yr}^{-1}$ for at least the final two subsequent observed snapshots, reflecting a heightened star formation rate for at least a $\sim500\ \mathrm{Myr}$ time period. 
If a galaxy temporarily rejuvenates for several snapshots but then returns to a passive state just before 1 Gyr (say), it is not considered rejuvenated at 1 Gyr, but may still be considered rejuvenated when measured against a shorter timescale.
This measure is calculated for central, satellite, and orphan galaxies. This ensures the galaxy has maintained significant star formation long enough to be otherwise indistinguishable from typical SF galaxies. Using this definition, we find around 10\% of typical PSB and passive galaxies rejuvenate within 1 Gyr of having a quiescent phase; this value agrees well with the 12\% found in MANGA observations by \cite{tanaka_hinotori_2024}. Similarly, the median mass of a rejuvenated galaxy is $10^{10.75} $\,M$_\odot$, closely matching the median seen in the MANGA population. 
\par

\begin{figure*}
  \includegraphics[width=2\columnwidth]{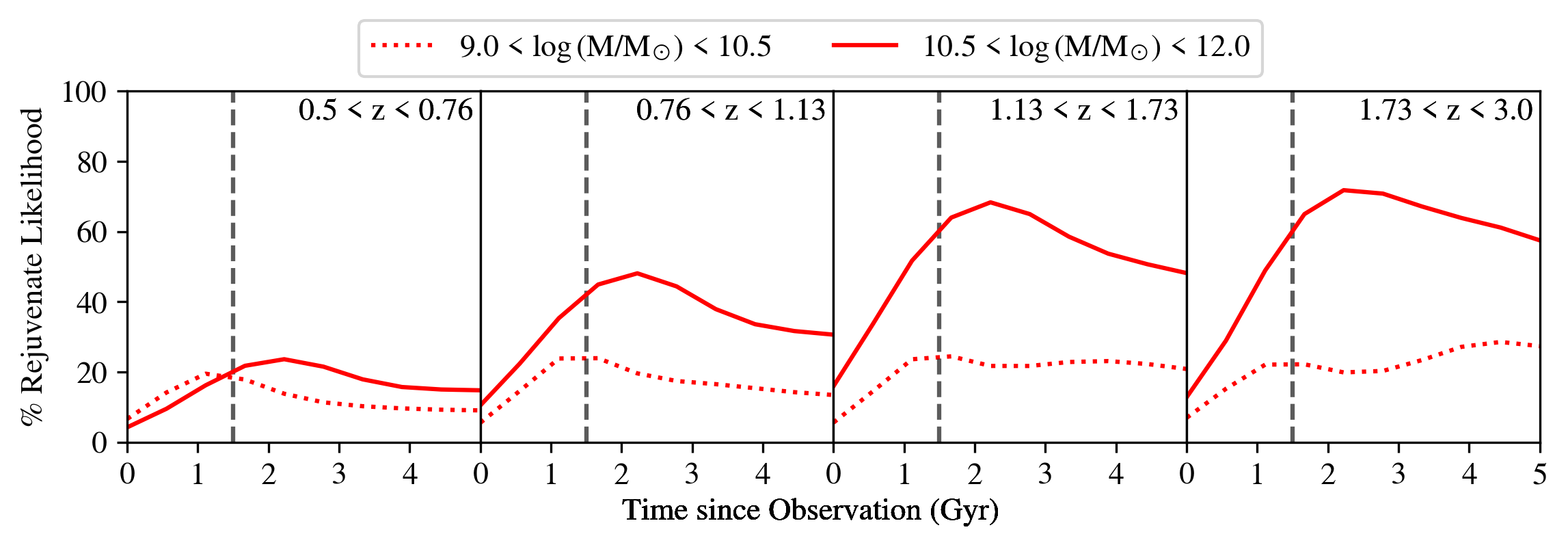}
  \caption{The measured rejuvenation rates for passive galaxies in the \Lgal{} mock dataset, as a function of time since observation, displayed in four redshift bins. 
  All passive galaxies in the indicated mass and redshift bin are examined at later times to see if they would then be classified as star-forming. As rejuvenation requires a galaxy to first be quiescent, only currently passive galaxies are tested. The vertical grey dashed line indicates a fixed time of 1.5~Gyr. The measured rejuvenation rates peak at $\sim 1.5 - 2$~Gyr after detection on the lightcone and then decrease with increasing time since the initial detection of the passive galaxies. This decrease is primarily because a rejuvenated SF galaxy often naturally re-quenches into a passive galaxy at later times, and thus would not be counted as a rejuvenated galaxy despite having passed through a rejuvenation phase.
  }
  \label{fig:psb_rejuv_rates}
\end{figure*}

The influence of rejuvenation in \Lgal{} can be seen in Figure \ref{fig:psb_rejuv_rates}, which shows the likelihood that a passive galaxy will have rejuvenated after a given amount of time. Both low and high mass galaxies have similar rejuvenation rates below $z=1$, peaking around 1.5 Gyrs after observation, with $\approx30\%$ of passive galaxies having been detected as a rejuvenated star-forming galaxy. However, at redshifts above $z=1$, high-mass galaxies show heightened rejuvenation rates of up to $\approx50\%$, while low-mass galaxies show a slight decrease in their overall rejuvenation rates compared to their low-redshift counterparts. Higher-mass galaxies in \Lgal{} are typically centrals; within \Lgal{} they can rejuvenate primarily through the accretion of new cold gas via gas-rich minor mergers, the impact of which is discussed in \cite{yates_dilution_2014}. Low-mass galaxies can rejuvenate via the above routes, but can also often rejuvenate through splashback events, where they temporarily become satellites of a larger sub-halo (preventing accretion of baryonic matter) and later return to becoming a central. 
\par

As time since detection on the lightcone increases, both high and low mass galaxies show reduced rejuvenation rates, primarily due to the fact that these galaxies re-quench over time back into the passive population. Although in theory there is potential for multiple rejuvenation events to occur within the same galaxy, in practice we observe that very few galaxies within our \Lgal{} sample experience multiple rejuvenation events. This is at odds with the results from both hydrodynamical simulations \citep{nelson_first_2018} and observations \citep{tanaka_hinotori_2024}, which see multiple rejuvenation events in the history of around 10\% of rejuvenated galaxies. 
This difference may simply result from the strict definition of rejuvenation adopted in this study, or perhaps from the differences in the AGN models used by \Lgal{} and TNG, with the \Lgal{} AGN model preventing any significant gas cooling once established. 
\par

\subsection{Visibility Timescales and Quiescent Buildup}\label{sec:visibility_mock}
\begin{figure}
  \includegraphics[width=\columnwidth]{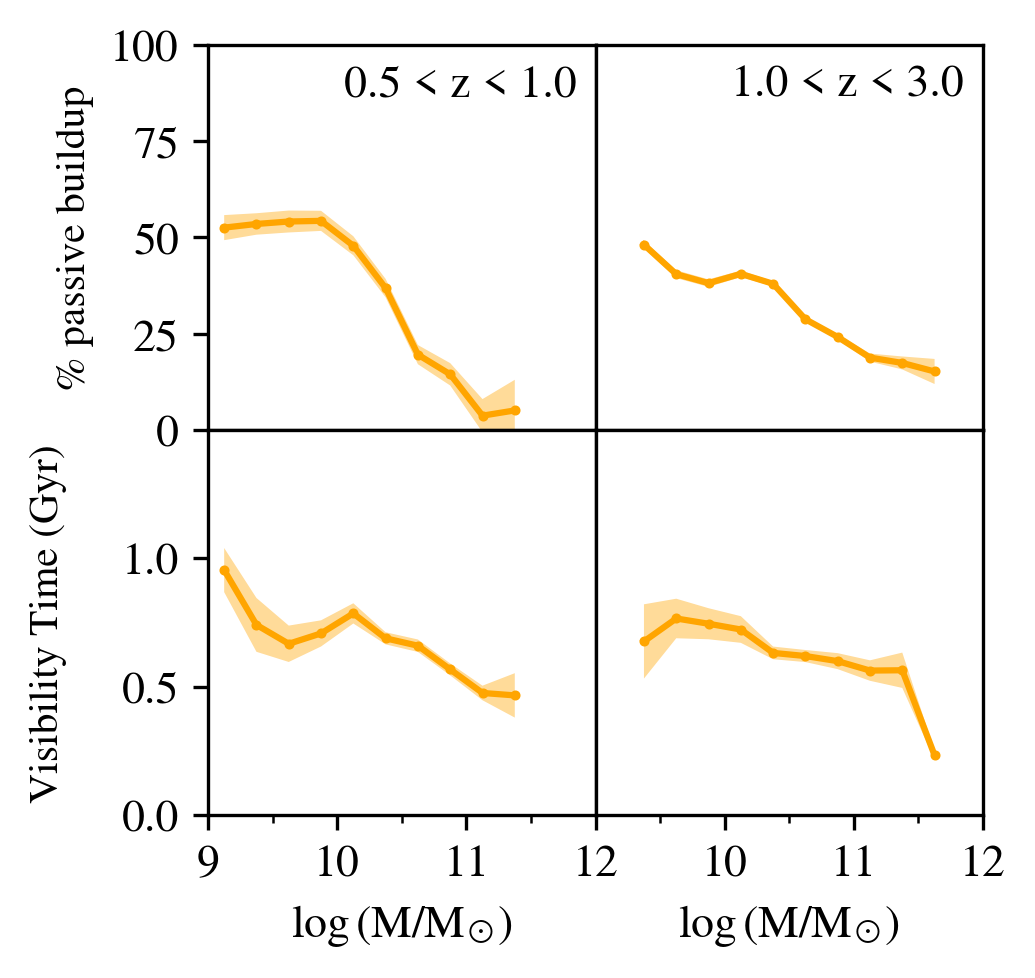}
  \caption{The percentage of passive galaxies that underwent a PSB phase in \Lgal{} (top panels) and the visibility time of the PSB phase (bottom panels). Percentages were calculated from the tracked SC history of each galaxy. The proportion of galaxies that underwent a PSB phase shows a strong mass dependence. The visibility timescale also appears to be dependent on stellar mass, with an overall average value of 0.64 Gyr.}
  \label{fig:mock_passive_buildup}
\end{figure}

By utilizing the full tracking of the \Lgal{} lightcone, we are able to directly count the number of snapshots a simulated galaxy exhibits PSB signatures in its SED, to obtain a PSB visibility timescale.
Applying this method across all galaxies, we find the mean PSB visibility time for simulated galaxies is 0.64 Gyr (Figure \ref{fig:mock_passive_buildup}, lower panel).
The average visibility times from \Lgal{} are consistent with the observational results of \cite{wild_star_2020}, who found visibility timescales in observations to be between 0.5 - 1 Gyr \citep[see also][]{belli_mosfire_2019}. However, previous studies have assumed 
a fixed visibility timescale across the available mass range. This is in contrast to what we observe directly from the \Lgal{} lightcone, which shows that galaxies at low stellar mass (M$_*$ < M$^{10}_{\odot}$) show visibility timescales closer to 1 Gyr. 
\par

The fully tracked properties of each galaxy in the \Lgal{} lightcone also allow us to directly calculate the number of quiescent galaxies that underwent a PSB phase. Doing so shows on average around 50\% of quiescent galaxies underwent a PSB phase at low stellar mass, with only an average of 20\% of galaxies undergoing a PSB phase at high stellar mass ($M_\ast$ > M$^{10.5}_{\odot}$), as shown in 
Figure \ref{fig:mock_passive_buildup} (upper panel).
While our calculated percentage of high-mass galaxies required to undergo a PSB phase are consistent with \cite{wild_star_2020} and \cite{belli_mosfire_2019}, the percentage of low-mass galaxies undergoing a PSB phase reflect the upper limits from both \cite{wild_star_2020} and \cite{belli_mosfire_2019}. 
\par

%% file: sections/observation_results.tex
\begin{figure*}
 \includegraphics[width=2\columnwidth]{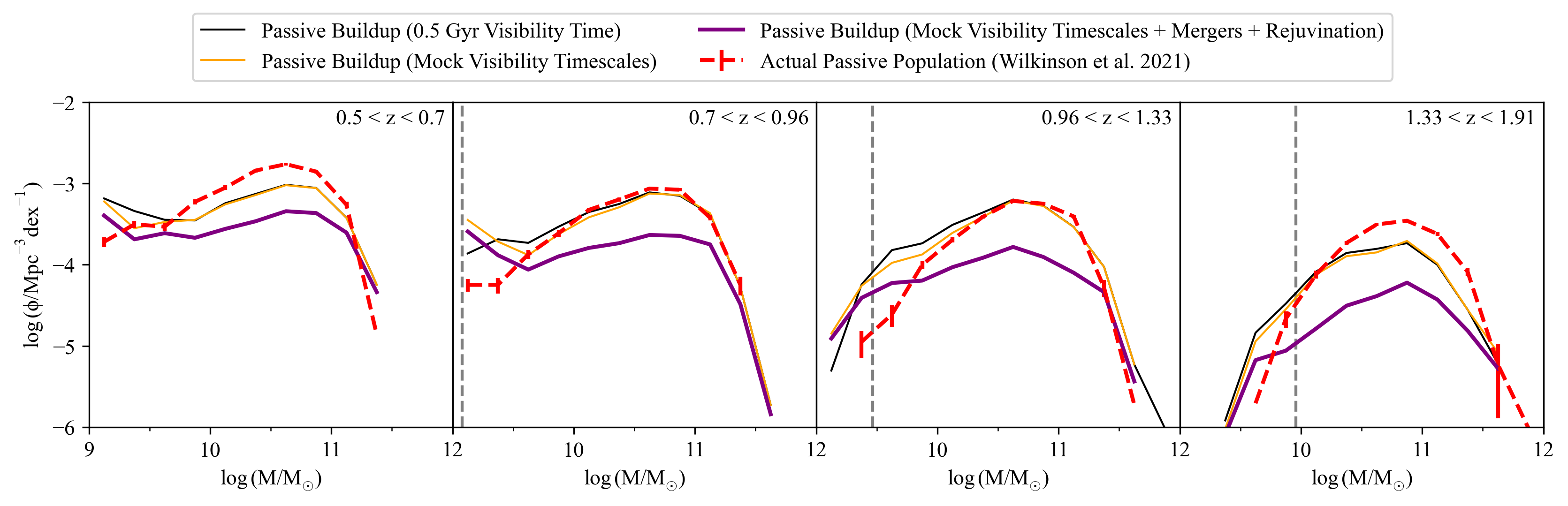}
 \caption{This diagram shows predictions for the build-up of the observed passive galaxy mass function in the UDS due to the contribution from PSBs, and the impact of additional constraints.
 The red dashed line represents the actual passive population observed in a given redshift bin, derived from the catalogue of \protect\cite{wilkinson_starburst_2021}. The black line demonstrates the predicted build-up from the previous bin assuming a fixed 0.5 Gyr lifetime for PSBs, obtained by combining the PSB and quiescent mass functions from the previous epoch.
 Alternative predictions using the mock visibility timescales are plotted with the orange line. The purple line shows the effects of taking into account mergers and rejuvenations between mass bins, which creates a significant drop in the predicted PSB contribution.
 The 95\% mass completeness limits for quiescent galaxies from the UDS are shown using the dashed gray line.
 }
 \label{fig:merge_rejuvination_smf}
\end{figure*}

Studies such as \cite{wild_star_2020} and \cite{belli_mosfire_2019} estimate the build-up of the quiescent population over time by combining the passive and post-starburst (PSB) populations at a given time and comparing them to the passive population at a later epoch. This method requires careful accounting for the fact that not all quiescent galaxies undergo a PSB phase, and that the PSB phase itself is transient in nature. Typically, a fixed visibility timescale for PSB features is assumed, with the visibility timescales expected to align with the estimated $0.5-1$ Gyr duration during which A- and F-class stars dominate a galaxy's SED. \cite{wild_star_2020} finds that between 25\% and 50\% of quiescent galaxies underwent a PSB phase in their sample at $z\sim 1$. \cite{belli_mosfire_2019} finds that approximately 50\% of the growth in the red sequence can be explained by rapid quenching at $z=2.2$, dropping to approximately 20\% by $z=1.4$. \cite{taylor_role_2023} find similar results, and also some evidence of potential overproduction of quiescent galaxies undergoing a PSB with a stellar mass below $M_*$ = $10^{10.7}$\,M$_\odot $. All of these studies assumed a fixed visibility timescale across the entire PSB population, typically 0.5 Gyr. However, we see from the results in Section \ref{sec:visibility_mock} that the typical PSB phase length in model galaxies varies with stellar mass, with the PSB phase of lower-mass quiescent galaxies taking closer to 1.0 Gyr.
\par

In Figure \ref{fig:merge_rejuvination_smf} we show the estimated passive buildup due to PSB galaxies in the UDS observations, using SMFs from the catalogue of \cite{wilkinson_starburst_2021}. The predicted passive buildup is plotted in a manner similar to \cite{wild_evolution_2016}, where the predicted quiescent stellar mass function for UDS galaxies at a given mass and redshift, $\Phi_{\mathrm{pas}}(M,z_0)$, is governed by the equation:
\begin{equation} \label{eq1}
\begin{split}
    \Phi_{\mathrm{pas}}(M,z_0)=&\left[f_\mathrm{safe,pas}(M,z_0,z_1)\Phi_{\mathrm{pas}}(M,z_1)\right]\\
    +&\frac{t_\mathrm{bin}(z_0,z_1)}{t_\mathrm{vis}(M,z_1)}\left[f_\mathrm{safe,PSB}(M,z_0,z_1)\Phi_{\mathrm{PSB}}(M,z_1)\right],
\end{split}
\end{equation}
where $\Phi_{\mathrm{pas}}(M,z_1)$ is the UDS stellar mass function in the previous redshift bin $z_1$. The two time values, $t_\mathrm{bin}(z_0,z_1)$ and $t_\mathrm{vis}(M,z_1)$, represent the length of time between the two redshift bins (e.g. 1.5 Gyr) and the expected visibility time of PSB features at a given mass and redshift respectively (e.g. fixed at 0.5 Gyr for \citealt{wild_evolution_2016}). Both $f_\mathrm{safe,pas}(M,z_0,z_1)$ and $f_\mathrm{safe,PSB}(M,z_0,z_1)$ represent the likelihood that a passive or PSB galaxy will \textit{not} merge or rejuvenate between two bins, which is calculated via the equation:
\begin{align}
    f_\mathrm{safe}=1-[(1-f_\mathrm{merge})(1-f_\mathrm{rejuvenate})],
\end{align}
where $f_\mathrm{merge}$ and $f_\mathrm{rejuvenate}$ represent the probability that a galaxy will merge or rejuvenate, respectively, which are derived for a given mass, class, and redshift from \Lgal{} using the method outlined in the previous sections. We show both the result of the typical fixed timescale and quiescent build-up used in works such as \cite{wild_evolution_2016} and \cite{belli_mosfire_2019}, as well as with the \Lgal{} estimated visibility timescales. We find that for galaxies in the UDS catalogue, when using a fixed visibility timescale of 0.5 Gyr, 
with no corrections for merging or rejuvenation, 
we require around 40\% of quiescent galaxies above $M_*$ = $10^{10.5}$\,M$_\odot$ to have undergone a PSB phase at $z=1$, in line with the upper range of the previously mentioned studies, but much higher than the $5-20\%$ we observe from \Lgal{} in Figure \ref{fig:mock_passive_buildup}. However, we also observe a significant overproduction of low-mass quiescent galaxies. We find this overproduction remains even when applying the longer visibility timescales seen for low-mass PSB galaxies from the \Lgal{} lightcone. Our \Lgal{} derived merger and rejuvenation rates show that many of these low-mass galaxies are likely to disappear via merging or rejuvenating between time bins. When applying the derived merger and rejuvenation rates we observe a significant reduction in the number density of quiescent galaxies predicted at all stellar masses (Figure \ref{fig:merge_rejuvination_smf}, purple curve).
At lower stellar mass, while the overproduction of the passive mass function is largely addressed by correcting for mergers and rejuvenation, we still find that a very large fraction of low-mass quiescent galaxies underwent a PSB phase.
\par

\begin{figure*}
 \includegraphics[width=2\columnwidth]{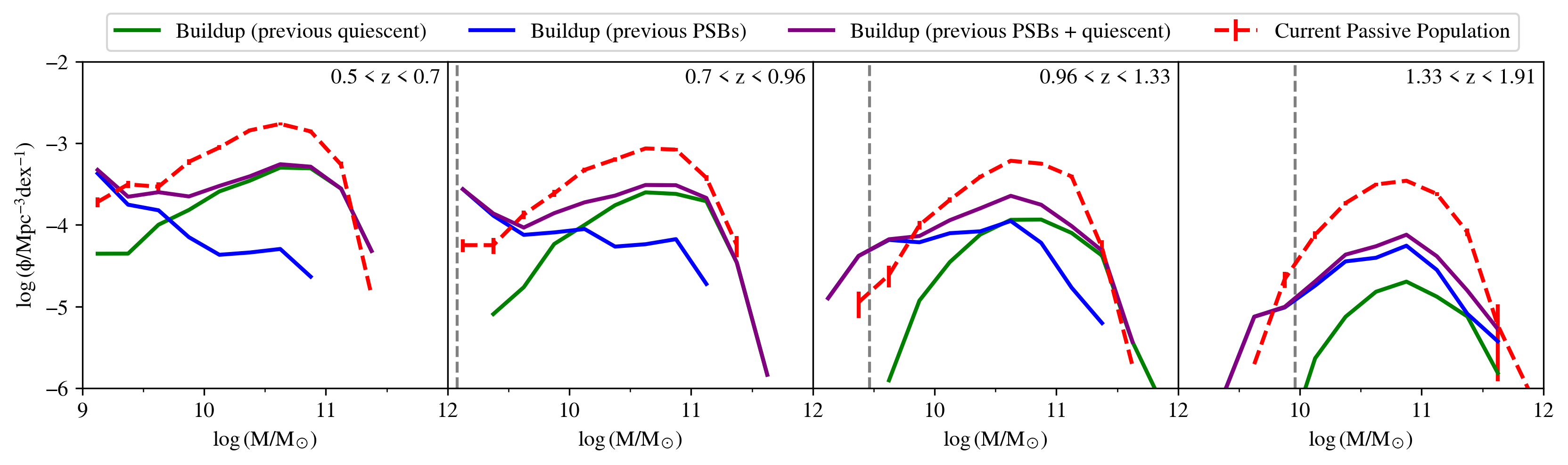}
 \caption{This diagram shows the various contributions to the build-up of the observed passive galaxy mass function, assuming the merger and rejuvenation model, and the visibility times from \Lgal{} (i.e., the purple curve from Figure 
 \ref{fig:merge_rejuvination_smf}). 
 The observed passive galaxy mass functions are shown with the red dashed line, with $1\sigma$ uncertainties. The contribution from passive galaxies from the previous epoch are shown in green, the PSB contribution in blue, with the summed contribution is shown in purple. The 95\% mass completeness limits for quiescent galaxies from the UDS are shown using the dashed gray line.}
 \label{fig:fit_passive_fraction}
\end{figure*}

In Figure \ref{fig:fit_passive_fraction}, we show the relative contribution of existing quiescent and PSB galaxies to the buildup of the observed
passive population, once accounting for 
the visibility timescales, merger rates, and rejuvenation rates derived from the 
\Lgal{} lightcone. Figure \ref{fig:fit_passive_fraction} effectively provides a more detailed examination of the model illustrated by the purple curve in Figure \ref{fig:merge_rejuvination_smf}.
The contribution of PSBs to the build-up of the passive mass function at various masses and redshifts is also shown in Table \ref{tab:fraction}. 
At all redshifts, we find that $60-80\%$ of the build-up in the low-mass passive mass function 
($M_*$ < $10^{10}$\,M$_\odot$) can be explained by PSBs (blue curve). At higher stellar mass ($M_*$ > $10^{10}$\,M$_\odot$), approximately $20-30\%$ of the passive build-up is due to PSBs at $z>1$, dropping to around $5\%$ by 
$z\sim 0.6$. These results largely agree with the values from \cite{belli_mosfire_2019}. Overall, we find that including 
the effects of visibility, merging, and rejuvenation has decreased the fraction of the high-mass build-up due to PSBs by approximately a factor of two in the UDS, but the fraction at low stellar masses remains very high compared to the observational estimates from \cite{belli_mosfire_2019}.
In theory, the very high fraction of low-mass quiescent galaxies explained by PSBs could be caused by \Lgal{} having inaccurate merger and rejuvenation rates when compared to observations. 
The rejuvenation rates for low-mass PSB galaxies may indeed be underestimated, with the results of \cite{kauffmann_quantitative_2014} and \cite{mintz_taking_2025} suggesting low-mass star-forming galaxies experience repeated cycles of starburst and quiescence, a rejuvenation behavior that is simply not modeled within the current version of \Lgal{}. When considering the merger rates within \Lgal{}, given that the rates for satellite galaxies are driven by the underlying DM simulation, in this case Millennium-I, and that the orphan merger rate is tuned to match this value, 
we consider it unlikely that the merger rates within \Lgal{} are drastically below the true values. Some of the overproduction we observe falls below the $M_* = 10^{9.5}\ \mathrm{M_\odot}$ mass limit defined in Section \ref{sec:supercolors}. \cite{guo_dwarf_2011} first highlighted an under-prediction of the number density of galaxies below this mass limit using \Lgal{} on the Millennium-I simulation (e.g. see figure 7). While this under-prediction will undoubtedly alter our results below $M_* = 10^{9.5}\ \mathrm{M_\odot}$, \cite{guo_dwarf_2011} also indicates that the expected galaxy number densities are altered by less than 0.1 dex above $M_* = 10^{9}\ \mathrm{M_\odot}$, of which the altered fraction that merge or rejuvenate will be even less significant with respect to our results. Nevertheless, there will be \textit{some} minor errors introduced to our results below $M_* = 10^{9.5}\ \mathrm{M_\odot}$. Another possibility is that some population of low-mass quiescent galaxies could have been missed or misclassified in the UDS catalogue, or the population of low-mass PSBs may be overestimated.
The reliability of the PCA classification method has been tested in \cite{wilkinson_starburst_2021} and \cite{maltby_identification_2016}, 
although notably not at very low stellar masses
($M_\ast$ < $10^{10}$\,M$_\odot$). Future spectroscopic and/or SED modeling \citep[e.g., see][]{wild_star_2020} may therefore be required to test the reliability of PCA selection at low stellar masses. Our results currently do not directly bin for galaxy environment, which is beyond the scope of the current work. \cite{chuter_galaxy_2011}, and later \cite{taylor_role_2023}, showed that low-mass passive and PSB galaxies preferentially reside in higher-density enviroments. Further exploration of our results as a function of environment may help to differentiate the conditions that preferentially lead to merging or rejuvenation, and when the \Lgal{} modeling begins to break down. Provisionally, however, we conclude that rapid quenching would appear to account for a significantly higher fraction of the low-mass quiescent population compared to high-mass galaxies.
\par

\begin{table*}
 \centering
 \begin{tabular}{cccccccc}
   Mass & & 0.5 < z $\le$ 0.7 & 0.7 < z $\le$ 0.96 & 0.96 < z $\le$ 1.33 & 1.33 < z $\le$ 1.91 & & 0.5 < z $\le$ 1.91\\
   \hline
  $10^{9}$\,M$_\odot$ $<$ $M_*$ $\le$ $10^{10}$\,M$_\odot$ & & 79.2 & 75.7 & 86.9 & 63.1& & 79.2\\
  $10^{10}$\,M$_\odot$ $<$ $M_*$ $\le$ $10^{12}$\,M$_\odot$ & & 4.71 & 15.0 & 27.8 & 17.7& & 20.6\\
 \end{tabular}
 \caption{The percentage of the build-up in the passive galaxy mass function that is due to PSBs, for various redshifts and in two mass ranges. These estimates are based on fitting to the observed build-up in the passive mass function (see Figure \ref{fig:fit_passive_fraction}), using visibility timescales, merger rates, and rejuvenation rates from the \Lgal{} lightcone.}
 \label{tab:fraction}
\end{table*}

%% file: sections/conclusion.tex
In this work, we used a mock observational lightcone that was generated using the SAM \Lgal{} to specifically replicate the properties of the UDS observational catalogue. The mock observational lightcone covers 1 square degree with a redshift range of $0.5<z<3$ and minimum mass limit of $M_{\ast}>10^{9.5}$\,M$_\odot$, containing over 100,000 simulated galaxies within its field-of-view. 
\par

We use the \Lgal{} lightcone to isolate the length of time PSB features are typically visible in a galaxy SED, and classified as such according to photometric PCA analysis. We also derive the merger and rejuvenation rates for each type of simulated galaxy in Section \ref{sec:merger_rejuvination_rates}, obtaining estimates for the fraction of PSB galaxies that will undergo a merger or rejuvenation over a range of timescales, masses, and redshifts.
\par

We apply these derived visibility timescales and merger and rejuvenation rates to more accurately model the predicted growth of the quiescent population when applied to observational data from the UDS catalogue of \cite{wilkinson_starburst_2021}. 
We find that $\sim 20-30\%$ of quiescent galaxies above $10^{10} $\,M$_\odot$ at $z>1$ underwent a PSB phase that was visible for 500 Myr, consistent with estimates from \cite{belli_mosfire_2019}.
We also find that the fraction of high-mass quiescent galaxies that undergo a PSB phase decreases towards low redshift, similar to the results of \cite{belli_mosfire_2019}, \cite{wild_star_2020}, and \cite{taylor_role_2023}.
Without corrections to the assumed visibility timescales to account for mergers and rejuvenation, the estimated fractional contribution from PSBs would be higher by approximately a factor of two, and comparable with earlier estimates from \cite{wild_evolution_2016}. At low stellar mass, however, 
the high space-density of PSBs can readily over-predict the build-up of the passive galaxy mass function without corrections for mergers, rejuvenation, and visibility timescales. With those corrections, we find that at least
$80\%$ of galaxies below $10^{10} $\,M$_\odot$ underwent a PSB phase in the UDS, 
assuming they are visible for longer than their high-mass counterparts ($\sim 750$ Myr on average), and we see that the remaining PSBs still readily over-predict the build-up of the passive galaxy mass function. This result lends further credence to the idea that low-mass star-forming galaxies experience repeated cycles of starburst as suggested in \cite{kauffmann_quantitative_2014} and \cite{mintz_taking_2025}.
\par

Overall, we conclude that mergers and rejuvenation can have a significant impact on the build-up of the passive galaxy mass functions.
Our results also provide further evidence that low-mass galaxies follow a distinctly different quenching pathway to that followed by high-mass galaxies. 
\par

%% file: sections/appendix.tex
The percentage of galaxies that will have merged after a given time period are shown in table \ref{tab:passive_merger_probablity_table}. Merger rates are calculated directly from \Lgal{} with Star forming, quiescent, and post starburst galaxies shown separately. Rejuvenation rates, also calculated directly using \Lgal{}, are shown in table \ref{tab:psb_rejuvination_probablity_table}.

\begin{table*}
\centering
\begin{tabular}{ l c c  c c c c llllllllll}
 & & &\multicolumn{4}{c}{Star Forming}&&\multicolumn{4}{c}{Passive}& &\multicolumn{4}{c}{Post Starburst}\\
 \hline
\multicolumn{2}{c}{}& &\multicolumn{4}{c}{Time/ Gyr} & & \multicolumn{4}{c}{Time/ Gyr}& & \multicolumn{4}{c}{Time/ Gyr}\\
\hline
\multicolumn{2}{c}{z}& Mass&  0.5& 1.0& 1.5& 2.0 & & 0.5& 1.0& 1.5&2.0  & & 0.5& 1.0& 1.5&2.0  \\
\hline
\multicolumn{2}{c}{ 0.5 < z < 1.0 }&  9.0 < \,M$_\odot$ < 9.5 &  0.4& 1.0& 1.4& 2.4& & 12.5& 23.4& 28.8&34.5 & & 6.1& 13.5& 19.1&24.2\\
\multicolumn{2}{c}{  }&  9.5 < \,M$_\odot$ < 10.0 &  0.9& 2.4& 2.7& 3.8& & 11.2& 21.0& 26.8&31.1 & & 1.4& 4.3& 6.5&15.8\\
\multicolumn{2}{c}{  }&  10.0 < \,M$_\odot$ < 11.0 &  1.5& 3.8& 4.5& 6.5& & 6.1& 11.2& 14.9&17.7 & & 0.0& 0.0& 0.0&0.0\\
\hline
\multicolumn{2}{c}{ 1.0 < z < 2.0 }&  9.0 < \,M$_\odot$ < 9.5 &  0.6& 1.4& 2.2& 3.6& & 19.4& 33.6& 42.2&50.3 & & 8.9& 19.9& 29.1&37.0\\
\multicolumn{2}{c}{  }&  9.5 < \,M$_\odot$ < 10.0 &  1.0& 2.4& 3.5& 5.3& & 18.4& 33.0& 41.1&48.3 & & 7.3& 21.1& 28.7&39.2\\
\multicolumn{2}{c}{  }&  10.0 < \,M$_\odot$ < 11.0 &  2.2& 4.1& 5.8& 7.8& & 9.3& 16.3& 21.2&26.3 & & 2.9& 9.7& 16.5&19.4\\
\hline
\multicolumn{2}{c}{ 2.0 < z < 3.0 }&  9.0 < \,M$_\odot$ < 9.5 &  0.3& 1.3& 2.9& 4.2& & 37.1& 52.9& 65.7&72.9 & & 0.0& 0.0& 26.7&53.3\\
\multicolumn{2}{c}{  }&  9.5 < \,M$_\odot$ < 10.0 &  0.9& 3.7& 6.3& 9.0& & 37.1& 52.9& 65.7&72.9 & & 18.7& 43.1& 58.9&68.5\\
\multicolumn{2}{c}{  }&  10.0 < \,M$_\odot$ < 11.0 &  1.4& 4.3& 6.7& 8.6& & 24.8& 44.6& 54.2&58.8 & & 8.8& 38.3& 63.3&66.2\\
\hline
\end{tabular}
    \caption{The percentage likelihood SF, passive and PSB galaxies will merge for a range of masses, redshifts, and timescales.}
    \label{tab:passive_merger_probablity_table}
\end{table*}

\begin{table*}
\centering
\begin{tabular}{ l c c  c c c c }
\multicolumn{2}{c}{}& &\multicolumn{4}{c}{Time/ Gyr}\\
\hline
\multicolumn{2}{c}{z}& Mass&  0.5& 1.0& 1.5& 2.0\\
\hline
\multicolumn{2}{c}{ 0.5 < z < 1.0 }&  9.0 < \,M$_\odot$ < 9.5 &  13.4& 25.1& 31.9& 32.8\\
\multicolumn{2}{c}{  }&  9.5 < \,M$_\odot$ < 10.0 &  11.9& 23.2& 30.6& 31.6\\
\multicolumn{2}{c}{  }&  10.0 < \,M$_\odot$ < 11.0 &  7.6& 17.5& 27.2& 29.5\\
\hline
\multicolumn{2}{c}{ 1.0 < z < 2.0 }&  9.0 < \,M$_\odot$ < 9.5 &  15.2& 29.0& 37.5& 38.7\\
\multicolumn{2}{c}{  }&  9.5 < \,M$_\odot$ < 10.0 &  14.3& 28.2& 37.6& 38.6\\
\multicolumn{2}{c}{  }&  10.0 < \,M$_\odot$ < 11.0 &  11.3& 23.2& 32.8& 35.1\\
\hline
\multicolumn{2}{c}{ 2.0 < z < 3.0 }&  9.0 < \,M$_\odot$ < 9.5 &  18.5& 39.2& 57.0& 60.5\\
\multicolumn{2}{c}{  }&  9.5 < \,M$_\odot$ < 10.0 &  17.8& 37.4& 53.4& 55.5\\
\multicolumn{2}{c}{  }&  10.0 < \,M$_\odot$ < 11.0 &  13.3& 28.4& 41.4& 44.0\\
\hline
\end{tabular}
    \caption{The percentage likelihood that a passive galaxy will rejuvenate for a range of masses, redshifts, and timescales.}
    \label{tab:psb_rejuvination_probablity_table}
\end{table*}